# Combination of Adomian decomposition method with Fourier transform for solving the squeezing flow influenced by a magnetic field

Mohammad Ramezani[1], Salman Nourazar[1*], Hamid Reza Dehghanpour[2]

1-Mechanical Engineering Department, Amirkabir University of Technology, Tehran, Iran

2-Physics Group, Tafresh University, Tafresh, Iran

Corresponding author, email: icp@aut.ac.ir

**Abstract:**

In this paper the Fourier transform combined with Adomian decomposition method (FTADM) is applied for solving the squeezed unsteady flow between parallel plates influenced by an inclined magnetic field. By moving these plates toward each other, the squeezing flow which is perpendicular to the plates is appeared. We assume that inclination varies from zero to ninety degrees. The momentum and energy equations are solved using the Adomian decomposition method with the combination of Fourier transform. The effect of the squeezed number, the angle of magnetic inclination, and the bottom plate suction/injection on the velocity and temperature are studied. The results show that by increasing the squeeze number, the intensity of the magnetic field and the magnetic inclined angle may increase the velocity near up and bottom plates in the longitudinal direction. However, the velocity near the midway of up and bottom plates may decrease. Moreover, our results show more accuracy and a smaller number of calculations when compared to the previous numerical simulations. This may be attributed to the fact that in the FTADM, one is able to incorporate all boundary conditions into the solution. However, in the semi-analytical methods, the solution may be accurate in a limited portion of the solution domain because only a part of boundary conditions is imposed into the solution.

**Key Words**: Squeezing, Fourier transform, Adomian decomposition method, Adomian polynomials, Non-linear differential equations

**Introduction:**

The study of squeezing flow between two infinite parallel plates has been very popular for researchers since the last decade. By moving these plates toward each other, the squeezing flow which is perpendicular to the plates appears. The squeezing flow has a vast majority of applications, such as power transmission, polymer processing, hydraulic lifts, cooling water, and so on. Many of these flows may be investigated by squeezing flow, for example, the lubrication problem may be considered as such flow [1]. Because of these applications, the study of momentum and heat transfer in squeezing flow has attracted a lot of attention. The primary study of squeezed flow between two parallel plates was performed by Stefan [1] where he studied the lubrication mechanism. This study was a start for more extensive works to understand the behavior of the squeezed flow between two planar plates. Ran et al. [2] used the homotopy analysis method (HAM) to investigate the squeezed flow between two planar plates, slowly moving towards each other. Mustafa et al. [3] reported characteristics of the heat and mass transfer in a viscous fluid squeezed between two infinite parallel plates. Sheikholeslami et al. [4] utilized neural network to estimate heat transfer rate for Nanofluid of alumina through a channel. Khan et al. [5] used the variation of parameters method (VPM) to study the behavior of copper nanoparticles between parallel plates. Dib et al. [6] employed Duan-Rach method to obtain a recursive convergent solution for momentum and energy equation for an unsteady squeezing problem of a Nanofluid flow. Hayat et al. [7] studied heat and mass transfer for electromagnetic squeezed flow through a Riga plate considering thermal radiation and chemical reaction. Hayat et al. [8] also considered the squeezed flow for Nanotubes of carbon as a Darcy-Forchheimer porous media. Ahmed et al. [9] carried out a theoretical study on characteristics of natural and forced convection in flow of Sutterby fluid in a squeezing flow. Investigation of momentum and heat transfer in squeezing flow influenced by a magnetic field has been very important for many researchers with a variety of applications such as MHD (Magneto-Hydro-Dynamic) generator, nuclear reactors, studies of plasma, geometrical extractions, and control of boundary layer [10-25]. Recently many researchers studied the behavior of the squeezed flow influenced by a magnetic field using different methods. Siddique et al. [26] studied 2-D flow of a viscous MHD fluid between parallel plates influenced by a magnetic field by using the homotopy perturbation method. Domairry and Aziz [27] performed a study on MHD squeezed flow between parallel disks given the suction/injection of the fluid. Haq et al. [28] studied MHD squeezing flow of a fluid through a sensor plate. Investigation of magnetohydrodynamic (MHD) squeezing flow of a nanofluid through a stretching porous media was performed by Hayat et al. [29]. The Adomian decomposition method (ADM) [30-37] is a well-known systematic method for the practical solution of linear, nonlinear, deterministic, and stochastic operator equations. These equations include ordinary differential equations (ODEs), partial differential equations (PDEs), integral equations, and integrodifferential equations. The ADM method is a powerful technique, which provides efficient algorithms for analytic approximate solutions for real-world applications in the applied sciences and engineering. A.M. Wazwaz studied the Bratu problem and Lane-Emden using the Adomian decomposition method [38-39]. J.S. Duan solved nonlinear fractional ordinary differential equations [40]. Y.T. Yang used the Adomian decomposition method for solving the periodic base temperature in convective longitudinal fins [41]. Recently, Nourazar et al. [42] developed a new method by a modification to the Adomian decomposition method using the Fourier transform. In the present work, we use the method of Nourazar et al. [42] to solve the momentum and energy equations of squeezing flow influenced by a magnetic field. Effects of different parameters such as: squeezing number, the intensity of the magnetic field, the bottom plate suction/injection parameter, and inclination angle on velocity and temperature profiles are studied. In using the semi-analytical methods, ADM, the convergence of the solution is not violated when the number of terms of series solution is increased. In other words, the convergence of the ADM is independent of the number of terms of series solution. Since the convergence of the ADM is verified. However, in the numerical methods the accuracy of the results, mesh refinement, is limited by convergency problems.

**The Adomian decomposition method (ADM)**

The Adomian decomposition method (ADM) is proposed by Adomian [43-45]. Assume the following equation:

$$G(u) = g(\eta), \tag{1}$$

where $G$ is an arbitrary operator that may be partitioned into the linear operator and a non-linear operator as:

$$G(u) = L(u) + N(u) = g(\eta). \tag{2}$$

The unknown function $u$ of the linear operator, $L$, may be written as a series solution:

$$u = \sum_{n=0}^{\infty} u_n, \tag{3}$$

where $n \geq 0$ and $u_n$ can be calculated by recursive equations:

$$L(u) = L\left(\sum_{n=0}^{\infty} u_n\right), \tag{4}$$

and non-linear operator $N$ can be written as:

$$N(u) = \sum_{n=0}^{\infty} A_n = \sum_{n=0}^{\infty} \frac{1}{n!} \frac{d^n}{d\lambda^n}\left[N\left(\sum_{n=0}^{\infty} \lambda^i u_i\right)\right]_{\lambda=0}, \tag{5}$$

and $N(u)$ may be rewritten as follows:

$$N(u) = N(u_0) + \left(\sum_{n=0}^{\infty} u_n\right) N'(u_0) + \frac{1}{2!}\left(\sum_{n=0}^{\infty} u_n\right)^2 N''(u_0) + \frac{1}{3!}\left(\sum_{n=0}^{\infty} u_n\right)^3 N'''(u_0) + \ldots$$

$$= N(u_0) + (u - u_0) N'(u_0) + \frac{1}{2!}(u - u_0)^2 N''(u_0) + \frac{1}{3!}(u - u_0)^3 N'''(u_0) + \ldots, \tag{6}$$

substituting Eq. (3) and (4) into Eq. (2):

$$L\left(\sum_{n=0}^{\infty} u_n\right) + \sum_{n=0}^{\infty} A_n = g(\eta), \tag{7}$$

where $u_n$ can be obtained by solving the recursive equations obtained from Eq. (7). Assume the overall error associated with series solution as follows [46]:

$$R(\eta) = L\left(\sum_{n=0}^{\infty} u_n\right) + \sum_{n=0}^{\infty} A_n - g(\eta). \tag{8}$$

Assuming that the overall error in connection with the series solution, Eq. (7), defined by Eq. (8) is square integrable as, $\text{Re}s = \int_{\Omega} R(\eta)^2 d\eta$. For a proper approximation $Res$ must tends to zero when the number of series terms approaches infinity. In our calculations the Duan and Rach [47] controlling parameter, $h$, for convergence is chosen to be equal to zero.

**Basic idea of FTADM**

Taking Fourier transform from Eq. (7) [48]:

$$L\left(\sum_{n=0}^{\infty} u_n\right) + \sum_{n=0}^{\infty} A_n = g(\omega), \quad (9)$$

expanding Eq. (9) as a recursive equation:

$$\sum_{n=1}^{\infty} L(u_i) + \sum_{n=0}^{\infty} A_i = 0, \quad (10)$$

the recursive equations, Eq. (10), could be written as:

$$L(u_k) + A_{k-1} = 0. \quad (11)$$

Where $u_0, u_1, u_2, u_3, ..., u_n$ are the components of series solution of $u = \sum_{n=0}^{\infty} u_n$ and Adomian polynomials are written as:

$$\sum_{k=0}^{\infty} A_k = \sum_{k=0}^{\infty} \frac{1}{k!} \frac{d^k}{d\lambda^k}\left[N\left(\sum_{k=0}^{\infty} \lambda^i u_i\right)\right]_{\lambda=0}, \quad (12)$$

## Governing equations

The problem to be considered is the unsteady squeezing flow of an incompressible electrically conducting fluid confined between two infinite parallel plates subjected to an inclined external magnetic field B (Fig.1). The x-axis is along the direction the lower plate of the channel and the y-axis is normal to it. The time-variable magnetic field $B = (B_m \cos\gamma, B_m \sin\gamma, 0)$, in which $B_m$ donates $B_0 = (1-\alpha t)^{-\frac{1}{2}}$, is applied at an inclination angle $\gamma$ with respect to the x-axis. The induced magnetic field is assumed to be negligible for a small magnetic Reynolds number. The gap between $H(t) = l(1-\alpha t)^{\frac{1}{2}}$, changes with the time t, where $l$ is the initial gap between the plates at the time $t = 0$ [49]. The case of $\alpha > 0$ corresponds to the squeezing motion of plates, as $\alpha > 0$ the plates move apart. Along the direction normal to the xy-plane, the velocity and temperature can be seen as unchanged. The Ohm's law gives the form of the current density vector J [50]:

$$J = \sigma(V \times B) = (0, 0, uB_m \sin\gamma - vB_m \cos\gamma) \quad (13)$$

in which $\sigma$ is the electrical conductivity, and $V = (u, v, 0)$ is the velocity vector utilizing eq.13, we obtain the Lorents force

$$J \times B = (\sigma B_m^2 v \sin\gamma \cos\gamma - \sigma B_m^2 u \sin^2\gamma, \sigma B_m^2 u \sin\gamma \cos\gamma - \sigma B_m^2 v \sin\gamma \cos^2\gamma, 0) \quad (14)$$

And the Joule heating

$$\frac{1}{\sigma} J.J = \sigma B_m^{\;2}(u^2 \sin^2 \gamma + v^2 \cos^2 \gamma - 2uv \sin \gamma \cos \gamma) \tag{15}$$

The conservation equations for mass, momentum and energy in the presence of the suction or injection across the stretching lower plate, viscous dissipation, and Joule heating may be written as:

$$\frac{\partial u}{\partial x} + \frac{\partial v}{\partial y} = 0 \tag{16}$$

$$\frac{\partial u}{\partial t} + u \frac{\partial u}{\partial x} + v \frac{\partial u}{\partial y} = -\frac{1}{\rho} \frac{\partial p}{\partial x} + \frac{\mu}{\rho}(\frac{\partial^2 u}{\partial x^2} + \frac{\partial^2 u}{\partial y^2}) + \frac{\sigma B_m^{\;2}}{\rho} \sin \gamma (v \cos \gamma - u \sin \gamma) \tag{17}$$

$$\frac{\partial v}{\partial t} + u \frac{\partial v}{\partial x} + v \frac{\partial v}{\partial y} = -\frac{1}{\rho} \frac{\partial p}{\partial y} + \frac{\mu}{\rho}(\frac{\partial^2 v}{\partial x^2} + \frac{\partial^2 v}{\partial y^2}) + \frac{\sigma B_m^{\;2}}{\rho} \cos \gamma (u \sin \gamma - v \cos \gamma) \tag{18}$$

$$\frac{\partial T}{\partial t} + u \frac{\partial T}{\partial x} + v \frac{\partial T}{\partial y} = \frac{k}{\rho C_p}(\frac{\partial^2 T}{\partial x^2} + \frac{\partial^2 T}{\partial y^2}) + \frac{\mu}{\rho C_p}\left[ 2(\frac{\partial u}{\partial x})^2 + 2(\frac{\partial v}{\partial y})^2 + (\frac{\partial u}{\partial y} + \frac{\partial v}{\partial x})^2 \right]$$
$$+ \frac{\sigma B_m^{\;2}}{\rho C_p}(u \sin \gamma - v \sin \gamma)^2 \tag{19}$$

Here u and v are the velocity components of the fluid along the x and the y directions, respectively. T is the temperature. μ, ρ, Cp and k denote the effective dynamic viscosity, the density, the specific heat capacity at constant pressure, and the thermal conductivity of fluids, respectively.

The following boundary conditions are imposed at the lower and upper plates:

$$u = u_s = \frac{bx}{1-\alpha t}, v = v_s = -\frac{v_0}{\sqrt{1-\alpha t}}, T = T_0 \text{ at } y = 0, \tag{20}$$

$$u = 0, v = v_H = \frac{dH}{dt} = -\frac{\alpha l}{2\sqrt{1-\alpha t}}, T = T_H = T_0 + \frac{T_0}{1-\alpha t} \text{ at } y = H(t) \tag{21}$$

Here $u_s$ is the stretching velocity of the lower plate, $v_c$ is the mass flux velocity through the lower plate, $v_H$ is the velocity of the upper plates, $T_0$ denotes the temperature at the upper-plate surface, $T_H$ is the temperature at the upper-plate surface.

The mathematical analysis of the problem is simplified by introducing the following change of variables:

$$\eta = \frac{y}{H(t)}, v = v_H f(\eta), u = \frac{-x}{H(t)} v_H f'(\eta), \theta(\eta) = \frac{T-T_0}{T_H - T_0} \tag{22}$$

By substituting (22) into Eqs. (16) – (19), the continuity equation is satisfied and then Eqs. (17)–(19) become:

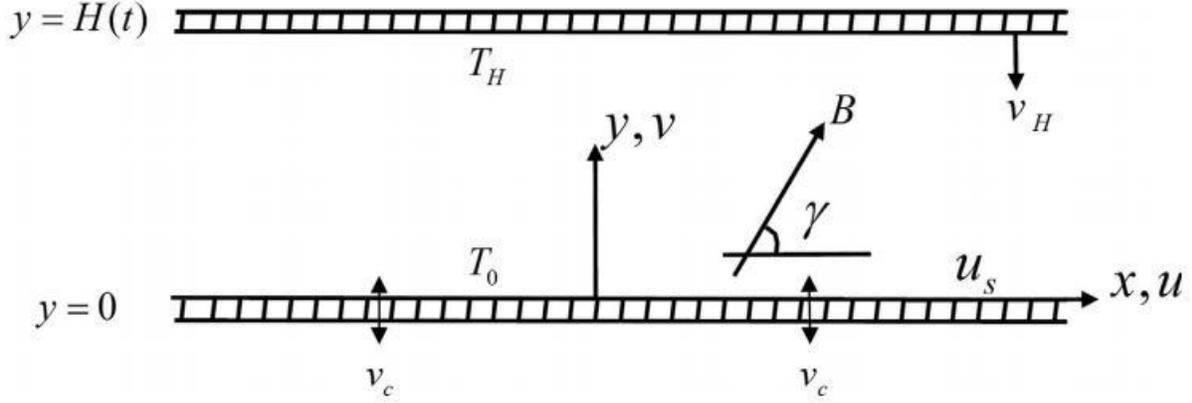

Fig. 1: Geometry configuration of the problem

The governing dimensionless momentum and energy equations for the squeezed flow influenced by a magnetic field can be written as [51]:

$$f^{iv} - S(\eta f''' + 3f'' + ff'' - ff''') - M^2 \sin\gamma(\sin\gamma f'' + 2\delta\cos\gamma f') = 0, \tag{23}$$

Where S is the squeeze number, $\eta$ is similarity coordinate, M is the magnetic parameter, $\gamma$ is inclination angle with respect to x axis and $\delta$ is the electrical conductivity.

$$\theta'' + \Pr S(f\theta' - \eta\theta' - 2\theta) + \Pr Ec\left(f''^2 + 4\delta^2 f'^2 + M^2(f'^2 \sin^2\gamma + f^2\delta^2\cos^2\gamma)\right)$$
$$+ 2\Pr Ec ff'\delta\cos\gamma\sin\gamma = 0, \tag{24}$$

boundary conditions are:

$$f(0) = S_b, \ f'(0) = R, \ f(1) = 1, \ f'(1) = 0 \tag{25}$$

$$\theta(0) = 0, \ \theta(1) = 1. \tag{26}$$

Where Pr is Prandtl number which is defined as the ratio of momentum boundary thickness to thermal boundary thickness and Eckert number, $Ec$, is defined as the ratio of mass transport by advective velocity ($u^2$, $u$ is the advective velocity) to the enthalpy difference of the plate and the fluid thickness ($C_p\Delta T$, $C_p$ is thermal conductivity coefficient at constant pressure and $T$ is temperature).

**Momentum equation:**

By taking the third and fourth order integral from Eq. (23) we may obtain $f''(0)$ and $f'''(0)$:

$$\int_0^1\int_0^\eta\int_0^\eta \left(f^{iv} - S(\eta f''' + 3f'' + ff'' - ff''') - M^2\sin\gamma(\sin\gamma f'' + 2\delta\cos\gamma f')\right)d\eta\,d\eta\,d\eta = 0, \tag{27}$$

$$\int_0^1 \int_0^\eta \int_0^\eta \int_0^\eta \left(f^{iv} - S(\eta f''' + 3f'' + ff'' - ff''') - M^2 \sin\gamma(\sin\gamma f'' + 2\delta\cos\gamma f')\right) d\eta\, d\eta\, d\eta\, d\eta = 0, \quad (28)$$

Taking Fourier transform of Eq. (23):

$$F\left(f^{iv} - S(\eta f''' + 3f'' + ff'' - ff''') - M^2 \sin\gamma(\sin\gamma f'' + 2\delta\cos\gamma f')\right) = 0, \quad (29)$$

$$F(f'(\eta)) = -f(0) + i\omega F(f(\eta)), \quad (30)$$

$$F(f''(\eta)) = -f'(0) - i\omega f(0) - \omega^2 F(f(\eta)), \quad (31)$$

$$F(f'''(\eta)) = -f''(0) - i\omega f'(0) + \omega^2 f(0) - i\omega^3 F(f(\eta)), \quad (32)$$

$$F\left(f^{iv}(\eta)\right) = -f'''(0) - i\omega f''(0) + \omega^2 f'(0) + i\omega^3 f(0) + \omega^4 F(f(\eta)), \quad (33)$$

$$f(\eta) = \sum_{n=0}^{\infty} f_n(\eta), \quad F\left(\sum_{n=0}^{\infty} f_n(\eta)\right) = \sum_{n=0}^{\infty} f_n(\omega), \quad (34)$$

$$\eta f''(\eta) = \sum_{n=0}^{\infty} A_n(\eta), \quad F\left(\sum_{n=0}^{\infty} A_n(\eta)\right) = \sum_{n=0}^{\infty} A_n(\omega), \quad (35)$$

$$ff'' + ff''' = \sum_{n=0}^{\infty} B_n(\eta), \quad F\left(\sum_{n=0}^{\infty} B_n(\eta)\right) = \sum_{n=0}^{\infty} B_n(\omega), \quad (36)$$

substituting Eqs. (30)-(33), $f''(0)$ and $f'''(0)$ into Eq. (29) and making use of Eqs. (34)-(36), Eq. (29) may be rewritten as:

(For more details see appendix)

$$\begin{aligned}
\sum_{n=0}^{\infty} f_n = f_0 &+ \left(-\frac{12}{\omega^4} + \frac{6i}{\omega^3}\right)\left(-S\left(\int_0^1\int_0^\eta\int_0^\eta\int_0^\eta \sum_{n=0}^{\infty} A_n d\eta d\eta d\eta d\eta + 3\int_0^1\int_0^\eta \sum_{n=0}^{\infty} f_n d\eta d\eta\right)\right) \\
&+ \left(-\frac{12}{\omega^4} + \frac{6i}{\omega^3}\right)\left(\int_0^1\int_0^\eta\int_0^\eta\int_0^\eta \sum_{n=0}^{\infty} B_n d\eta d\eta d\eta d\eta\right) + \left(-\frac{12}{\omega^4} + \frac{6i}{\omega^3}\right)\left(-M^2\sin\gamma\left(\sin\gamma\int_0^1\int_0^\eta \sum_{n=0}^{\infty} f_n d\eta d\eta\right)\right) \\
&+ \left(-\frac{12}{\omega^4} + \frac{6i}{\omega^3}\right)\left(2\delta\cos\gamma\int_0^1\int_0^\eta\int_0^\eta \sum_{n=0}^{\infty} f_n d\eta d\eta d\eta\right) + \left(\frac{6}{\omega^4} - \frac{2i}{\omega^3}\right)\left(-S\int_0^1\int_0^\eta\int_0^\eta \sum_{n=0}^{\infty} \widehat{A_n d\eta d\eta d\eta}\right) \\
&+ \left(\frac{6}{\omega^4} - \frac{2i}{\omega^3}\right)\left(-S\left(3\int_0^1 \sum_{n=0}^{\infty} \widehat{f_n d\eta} + \int_0^1\int_0^\eta\int_0^\eta \sum_{n=0}^{\infty} \widehat{B_n d\eta d\eta d\eta}\right)\right) + \left(\frac{6}{\omega^4} - \frac{2i}{\omega^3}\right)\left(-M^2\sin\gamma\left(\sin\gamma\int_0^1 \sum_{n=0}^{\infty} \widehat{f_n d\eta}\right)\right) \\
&+ \left(\frac{6}{\omega^4} - \frac{2i}{\omega^3}\right)\left(2\delta\cos\gamma\int_0^1\int_0^\eta \sum_{n=0}^{\infty} \widehat{f_n d\eta d\eta}\right) + \frac{S}{\omega^4}\left(\sum_{n=0}^{\infty} \widehat{A_n} - 3\omega^2 \sum_{n=0}^{\infty} \widehat{f_n} + \sum_{n=0}^{\infty} \widehat{B_n}\right) \\
&+ \frac{M^2\sin\gamma}{\omega^4}\left(-\omega^2\sin\gamma \sum_{n=0}^{\infty} \widehat{f_n} + 2i\omega\delta\cos\gamma \sum_{n=0}^{\infty} \widehat{f_n}\right),
\end{aligned} \quad (37)$$

the recursive equations for momentum equation may be written as:

$$f_k = (-\frac{12}{\omega^4} + \frac{6i}{\omega^3})\left(-S\left(\int_0^1\int_0^\eta\int_0^\eta\int_0^\eta A_{k-1}d\eta d\eta d\eta d\eta + 3\int_0^1\int_0^\eta f_{k-1}d\eta d\eta\right)\right)$$

$$+(-\frac{12}{\omega^4} + \frac{6i}{\omega^3})\left(\int_0^1\int_0^\eta\int_0^\eta\int_0^\eta B_{k-1}d\eta d\eta d\eta d\eta\right) + (-\frac{12}{\omega^4} + \frac{6i}{\omega^3})\left(-M^2\sin\gamma(\sin\gamma\int_0^1\int_0^\eta f_{k-1}d\eta d\eta)\right)$$

$$+(-\frac{12}{\omega^4} + \frac{6i}{\omega^3})\left(2\delta\cos\gamma\int_0^1\int_0^\eta\int_0^\eta f_{k-1}d\eta d\eta d\eta\right) + (\frac{6}{\omega^4} - \frac{2i}{\omega^3})\left(-S\int_0^1\int_0^\eta\int_0^\eta A_{k-1}d\eta d\eta d\eta\right)$$

$$(\frac{6}{\omega^4} - \frac{2i}{\omega^3})\left(-S(3\int_0^1 \widehat{f_{k-1}d\eta} + \int_0^1\int_0^\eta\int_0^\eta \sum_{n=0}^\infty \widehat{B_{k-1}d\eta d\eta d\eta})\right) + (\frac{6}{\omega^4} - \frac{2i}{\omega^3})\left(-M^2\sin\gamma(\sin\gamma\int_0^1 \widehat{f_{k-1}d\eta})\right) \quad (38)$$

$$+(\frac{6}{\omega^4} - \frac{2i}{\omega^3})\left(2\delta\cos\gamma\int_0^1\int_0^\eta \widehat{f_{k-1}d\eta d\eta}\right) + \frac{S}{\omega^4}\left(\widehat{A_{k-1} - 3\omega^2 f_{k-1} + B_{k-1}}\right)$$

$$+\frac{M^2\sin\gamma}{\omega^4}\left(-\omega^2\sin\gamma \widehat{f_{k-1}} + 2i\omega\delta\cos\gamma \widehat{f_{k-1}}\right)$$

$$f_k = F^{-1}(\widehat{f_k}),$$

which $A_0, A_1, A_2, ..., A_{k-1}, B_0, B_1, B_2, ...., B_{k-1}$ are the Fourier transforms of the Adomian polynomials.

$$f_0(\eta) = (R + 2S_b - 2)\eta^3 + (3 - 3S_b - 2R)\eta^2 + R\eta + S_b,$$

$$f_1(\eta) = \left(\frac{SR^2}{70} + \frac{2SRS_b}{35} - \frac{2SR}{35} + \frac{2SS_b^2}{35} - \frac{4SS_b}{35} + \frac{2S}{35}\right)\eta^7$$

$$+\left(\frac{\delta\sin(2\gamma)M^2R}{120} + \frac{\delta\sin(2\gamma)M^2S_b}{120} + \frac{\delta\sin(2\gamma)M^2}{60} - \frac{SR^2}{15} - \frac{7SRS_b}{30} + \frac{7SR}{30} - \frac{S}{5}\right)\eta^6$$

$$+\left(\frac{\sigma_2}{20} - \frac{SR}{5} - \frac{SS_b}{5} - \frac{S}{10} + \frac{M^2R}{40} + \frac{M^2S_b}{20} + \frac{2SR^2}{15} + \frac{3SS_b}{10} - \frac{M^2}{20} + \frac{2SRS_b}{5}\right)\eta^5$$

$$+\left(\frac{3S}{4} - \frac{SR}{4} - \frac{SS_b}{4} - \frac{\sigma_2}{8} - \frac{M^2R}{12} - \frac{M^2S_b}{8} - \frac{SR^2}{6} - \frac{SS_b^2}{2} + \frac{M^2}{8} - \frac{SRS_b}{2} + \frac{\sigma_4}{12} + \frac{\sigma_1}{8} + \frac{\sigma_3}{24}\right)\eta^4$$

$$+\left(\begin{array}{c}\frac{19SR}{42} - \frac{24S}{35} + \frac{SS_b}{14} + \frac{\sigma_2}{10} + \frac{11M^2R}{120} + \frac{M^2S_b}{10} + \frac{9SR^2}{70} + \frac{43SS_b^2}{70} \\ -\frac{M^2}{10} + \frac{47SRS_b}{105} - \frac{11M^2R\cos(2\gamma)}{120} - \frac{\sigma_1}{10} - \frac{\sigma_2}{12} - \frac{\sigma_3}{60} + \frac{\sigma_6}{12}\end{array}\right)\eta^3$$

$$+\left(\begin{array}{c}\frac{5S}{28} - \frac{5SR}{28} + \frac{13SS_b}{140} - \frac{\sigma_2}{40} - \frac{M^2R}{30} - \frac{M^2S_b}{40} - \frac{3SR^2}{70} \\ -\frac{19SS_b}{70} + \frac{M^2}{40} - \frac{6SRS_b}{35} + \frac{\sigma_4}{30} + \frac{\sigma_1}{40} + \frac{\sigma_5}{20} + \frac{\sigma_6}{20}\end{array}\right)\eta^2,$$

$$\sigma_1 = M^2S_b\cos(2\gamma)$$

$$\sigma_2 = M^2 \cos(2\gamma)$$

$$\sigma_3 = M^2 R \delta \sin(2\gamma)$$

$$\sigma_4 = M^2 R \cos(2\gamma)$$

$$\sigma_5 = M^2 \delta \sin(2\gamma)$$

$$\sigma_6 = M^2 S_b \delta \sin(2\gamma). \tag{39}$$

**Energy equation:**

$$\theta'' + \Pr S(f\theta' - \eta\theta' - 2\theta) + \Pr Ec\left(f''^2 + 4\delta^2 f'^2 + M^2(f'^2 \sin^2\gamma + f^2\delta^2 \cos^2\gamma + 2ff'\delta\cos\gamma\sin\gamma)\right) = 0, \tag{24}$$

$$\theta(0) = 0, \quad \theta(1) = 1, \tag{25}$$

taking double integral from Eq. (14):

$$\int_0^1 \int_0^\eta \begin{pmatrix} \theta'' + \Pr S(f\theta' - \eta\theta' - 2\theta) \\ + \Pr Ec\left(f''^2 + 4\delta^2 f'^2 + M^2(f'^2 \sin^2\gamma + f^2\delta^2 \cos^2\gamma + 2ff'\delta\cos\gamma\sin\gamma)\right) \end{pmatrix} d\eta d\eta = 0, \tag{40}$$

$$\theta'(0) = 1 + \Pr S \int_0^1 \int_0^\eta (f\theta' - \eta\theta' - 2\theta) d\eta d\eta + \Pr Ec \int_0^1 \int_0^\eta (f''^2 + 4\delta^2 f'^2) d\eta d\eta$$
$$+ \Pr Ec \int_0^1 \int_0^\eta M^2 (f'^2 \sin^2\gamma + f^2\delta^2 \cos^2\gamma + 2ff'\delta\cos\gamma\sin\gamma) d\eta d\eta. \tag{41}$$

Taking Fourier transform of Eq. (24):

$$F\begin{pmatrix} \theta'' + \Pr S(f\theta' - \eta\theta' - 2\theta) \\ + \Pr Ec\left(f''^2 + 4\delta^2 f'^2 + M^2(f'^2 \sin^2\gamma + f^2\delta^2 \cos^2\gamma)\right) \\ + 2\Pr Ec ff'\delta\cos\gamma\sin\gamma \end{pmatrix} = 0, \tag{42}$$

$$F(\theta'(\eta)) = -\theta(0) + i\omega F(\theta(\eta)), \tag{43}$$

$$F(\theta''(\eta)) = -\theta'(0) - i\omega\theta(0) - \omega^2 F(\theta(\eta)), \tag{44}$$

Substituting Eqs. (43) and (44) into Eq. (42):

$$F(\theta(\eta)) = \frac{1}{\omega^2} \begin{pmatrix} -\theta'(0) + \Pr S\left(F(f\theta' - \eta\theta' - 2\theta)\right) + \Pr Ec\left(F(f''^2) + 4\delta^2 F(f'^2)\right) \\ + \Pr Ec M^2\left(F(f'^2)\sin^2\gamma + F(f^2)\delta^2 \cos^2\gamma + 2F(ff')\delta\cos\gamma\sin\gamma\right) \end{pmatrix}, \tag{45}$$

substituting Eq. (41) into Eq. (45):

$$F(\theta(\eta)) = -\frac{1}{\omega^2}$$

$$-\frac{\Pr}{\omega^2}\begin{pmatrix} S\int_0^1\int_0^\eta (f\,\theta' - \eta\theta' - 2\theta)d\eta d\eta \\ +Ec\int_0^1\int_0^\eta (f''^2 + 4\delta^2 f'^2 + M^2(f'^2\sin^2\gamma + f^2\delta^2\cos^2\gamma + 2ff'\delta\cos\gamma\sin\gamma))d\eta d\eta \\ -S\left(F(f\,\theta' - \eta\theta' - 2\theta)\right) \\ -Ec\left(F(f''^2) + 4\delta^2 F(f'^2) + M^2(F(f'^2)\sin^2\gamma + F(f^2)\delta^2\cos^2\gamma + 2F(ff')\delta\cos\gamma\sin\gamma)\right) \end{pmatrix}, \quad (46)$$

$$\theta = \sum_{n=0}^\infty \theta_n(\eta), \quad F\left(\sum_{n=0}^\infty \theta_n(\eta)\right) = \sum_{n=0}^\infty \theta_n(\omega), \quad (47)$$

$$f\theta' - \eta\theta' - 2\theta = \sum_{n=0}^\infty C_n(\eta), \quad F\left(\sum_{n=0}^\infty C_n(\eta)\right) = \sum_{n=0}^\infty C_n(\omega), \quad (48)$$

$$f''^2 = \sum_{n=0}^\infty E_n(\eta), \quad F\left(\sum_{n=0}^\infty E_n(\eta)\right) = \sum_{n=0}^\infty E_n(\omega), \quad (49)$$

$$f'^2 = \sum_{n=0}^\infty G_n(\eta), \quad F\left(\sum_{n=0}^\infty G_n(\eta)\right) = \sum_{n=0}^\infty G_n(\omega), \quad (50)$$

$$f^2 = \sum_{n=0}^\infty H_n(\eta), \quad F\left(\sum_{n=0}^\infty H_n(\eta)\right) = \sum_{n=0}^\infty H_n(\omega), \quad (51)$$

$$ff' = \sum_{n=0}^\infty K_n(\eta), \quad F\left(\sum_{n=0}^\infty K_n(\eta)\right) = \sum_{n=0}^\infty K_n(\omega), \quad (52)$$

$$f''^2 + 4\delta^2 f'^2 + M^2(f'^2\sin^2\gamma + f^2\delta^2\cos^2\gamma + 2ff'\delta\cos\gamma\sin\gamma) = P(\eta)$$
$$P(\eta) = \sum_{n=0}^\infty P_n(\eta), \quad (53)$$

$$\sum_{n=0}^\infty E_n(\eta) + 4\delta^2 \sum_{n=0}^\infty G_n(\eta) + M^2\left(\sum_{n=0}^\infty G_n(\eta)\sin^2\gamma + \sum_{n=0}^\infty H_n(\eta)\delta^2\cos^2\gamma\right.$$
$$\left. +2\sum_{n=0}^\infty K_n(\eta)\delta\cos\gamma\sin\gamma\right) = \sum_{n=0}^\infty P_n(\eta), \quad (54)$$

$$\sum_{n=0}^\infty E_n(\omega) + 4\delta^2 \sum_{n=0}^\infty G_n(\omega) + M^2\left(\sum_{n=0}^\infty G_n(\omega)\sin^2\gamma + \sum_{n=0}^\infty H_n(\omega)\delta^2\cos^2\gamma\right.$$
$$\left. +2\sum_{n=0}^\infty K_n(\omega)\delta\cos\gamma\sin\gamma\right) = \sum_{n=0}^\infty P_n(\omega). \quad (55)$$

substituting Eqs. (47)-(55) into Eq. (46):

$$\sum_{n=0}^{\infty} \theta_n(\omega) = -\frac{1}{\omega^2}$$
$$-\frac{\Pr}{\omega^2}\left(\begin{array}{c} S\left(\int_0^1\int_0^\eta \sum_{n=0}^{\infty} C_n(\eta)d\eta d\eta - \sum_{n=0}^{\infty} C_n(\omega)\right) \\ Ec\left(\int_0^1\int_0^\eta \sum_{n=0}^{\infty} P_n(\eta)d\eta d\eta - \sum_{n=0}^{\infty} P_n(\omega)\right) \end{array}\right), \tag{56}$$

the recursive equations for energy equation may be written as:

$$\theta_0(\omega) = -\frac{1}{\omega^2}, \theta_0 = F^{-1}(\theta_0),$$

$$\theta_1(\omega) = -\frac{\Pr}{\omega^2}\left[S\left(\int_0^1\int_0^\eta C_0(\eta)d\eta d\eta - C_0(\omega)\right)\right] + \left[Ec\left(\int_0^1\int_0^\eta P_0(\eta)d\eta d\eta - P_0(\omega)\right)\right], \theta_1 = F^{-1}(\theta_1),$$

$$\theta_2(\omega) = -\frac{\Pr}{\omega^2}\left[S\left(\int_0^1\int_0^\eta C_1(\eta)d\eta d\eta - C_1(\omega)\right)\right] + \left[Ec\left(\int_0^1\int_0^\eta P_1(\eta)d\eta d\eta - P_1(\omega)\right)\right], \theta_2 = F^{-1}(\theta_2),$$

and so on

.
.
.

$$\theta_k(\omega) = -\frac{\Pr}{\omega^2}\left[S\left(\int_0^1\int_0^\eta C_{k-1}(\eta)d\eta d\eta - C_{k-1}(\omega)\right)\right] + \left[Ec\left(\int_0^1\int_0^\eta P_{k-1}(\eta)d\eta d\eta - P_{k-1}(\omega)\right)\right] \tag{57}$$
$$\theta_k = F^{-1}(\theta_k).$$

Which $C_0, C_1, C_2, ..., C_{k-1}, P_0, P_1, P_2, ...., P_{k-1}$ are the Fourier transforms of the Adomian polynomials respectively. The overall error for momentum and energy equations may be written as:

$$R\_1 = f^{iv} - S(\eta f''' + 3f'' + ff'' - ff''') - M^2\sin\gamma(\sin\gamma f'' + 2\delta\cos\gamma f'), \tag{58}$$

$$R\_2 = \theta'' + \Pr S(f\theta' - \eta\theta' - 2\theta) + \Pr Ec\left(f''^2 + 4\delta^2 f'^2 + M^2(f'^2\sin^2\gamma + f^2\delta^2\cos^2\gamma)\right)$$
$$+ 2\Pr Ec ff'\delta\cos\gamma\sin\gamma, \tag{59}$$

which R_1 and R_2 are overall errors for momentum and energy equations respectively.

**Results and discussion:** In the present work we use the method of Nourazar et al. [42] to solve the momentum and energy equations of squeezing flow influenced by a magnetic field. Effects of different parameters such as: squeezing number, intensity of magnetic field, the bottom plate suction/injection parameter and inclination angle on velocity and temperature profiles are studied. Figs.2 and 3 show the variations of overall error for momentum and energy equations versus the number of recursive terms in the FTADM recursive equations, Eq. (58) and (81) respectively. As may be seen from Figs. 2 and 3, once the number of recursive terms is increased the error associated with results decreases monotonically.

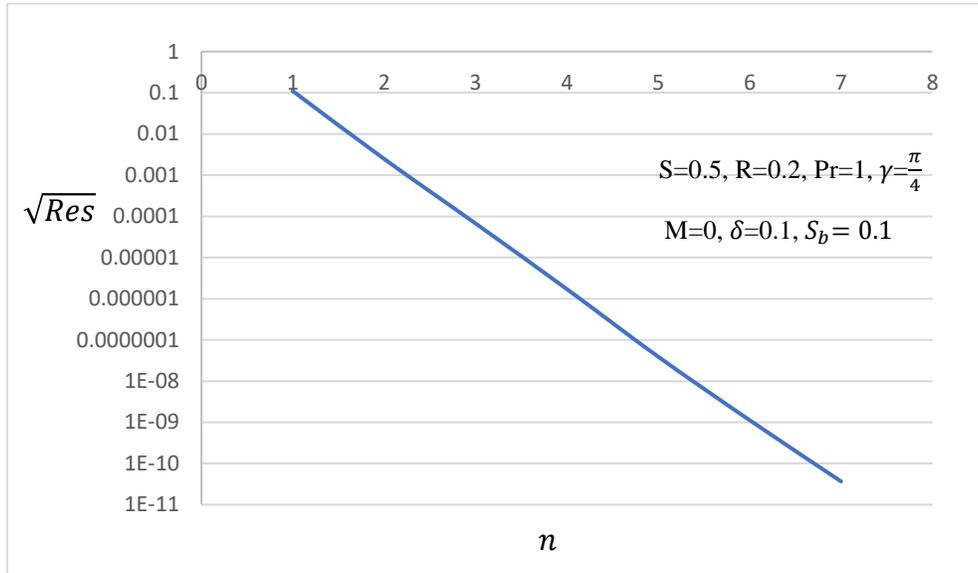

Fig. 2: variations of overall error of momentum equation versus the number of terms in recursive equation, $n=1,2,3,4,5,6,7$.

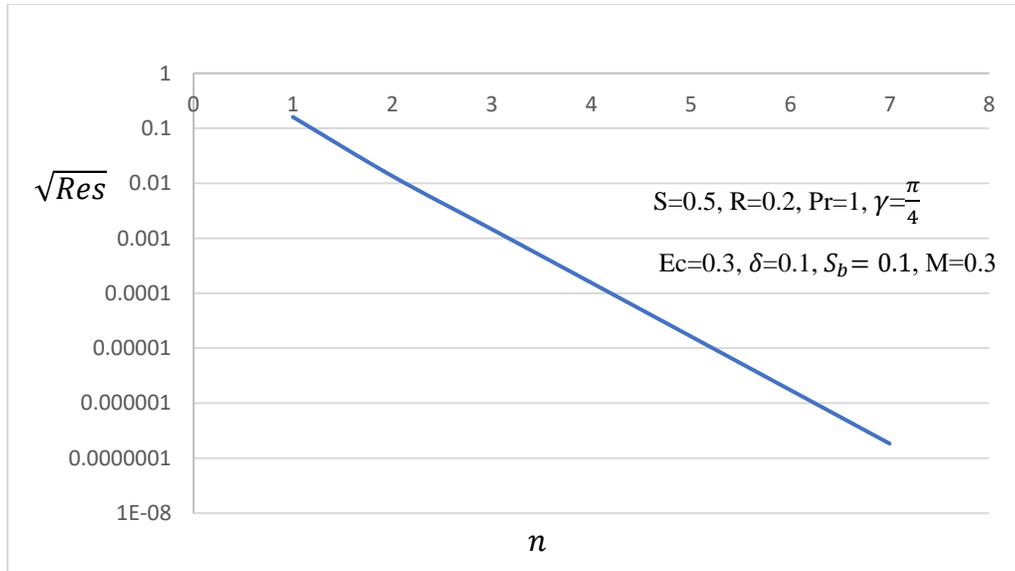

Fig. 3: variations of overall error of energy equation versus the number of terms in recursive equation, $n=1,2,3,4,5,6,7$.

Figs. 4 and 5 show the effect of the intensity of the magnetic field ($M$) on the velocity $f'(\eta)$ and temperature in the longitudinal direction. Fig. 4 shows that increasing the intensity of the magnetic field results in increasing the velocity profile in the longitudinal direction near the up and bottom plates however, the velocity near in the middle of up and bottom plates decreases slightly. This may be attributed to the intensity of the magnetic field. Fig. 5 shows that the temperature of fluid rises as the magnitude of magnetic intensity field $(M)$ increases and this increase is monotonic for small values of $M$, but a maximum value of temperature occurs in central region between up and bottom plates for large values of $M$. Figs. 6 and 7 show the effect of magnetic inclined angle $\gamma$ on velocity $f'(\eta)$ and temperature $\theta(\eta)$ in the longitudinal direction. We may see the same behavior for velocity and temperature profile as is seen in Figs. 4 and 5 because of the intensity of the magnetic field $(M)$. Figs. 8 and 9 show the influence of bottom-plate stretching parameter $(R)$ on velocity $f'(\eta)$ and temperature $\theta(\eta)$ in the longitudinal direction. Fig. 8 reveals that increasing of $R$ results in increasing the velocity near the bottom plate where the velocity near the upper plate decreases. By increasing the stretching parameter $(R)$, maximum velocity occurring in central region between up and bottom plates occurs on surface of the bottom plate. We conclude from Fig. 9 that increasing of $R$ the temperature decreases initially but increases subsequently. Figs. 10 and 11 show the effect of squeeze number $S$ on velocity $f'(\eta)$ and temperature $\theta(\eta)$ in the longitudinal direction. Fig. 10 shows that increasing of squeeze number results in decreasing of velocity of the fluid in the longitudinal direction near the up and bottom plates, but there is an opposite behavior for velocity in central region between up and bottom plates. Fig. 11 reveals that by increasing the value of $S$, the temperature decreases. Figs. 12 and 13 are showing the effect of bottom-plate suction/injection parameter $S_b$ on velocity $f'(\eta)$ and temperature $\theta(\eta)$ in the longitudinal direction. Fig. 12 shows that increase of $S_b$ induces the decrease in maximum velocity in the longitudinal direction. The temperature profile shows a decrease while the $S_b$ increases.

**Conclusion**: In the present study, we propose a new method using the Fourier transform combined with the Adomian decomposition method (FTADM) for solving the squeezed unsteady flow between parallel plates influenced by an inclined magnetic field. The influence of intensity of the magnetic field, the magnetic inclined angle, the bottom-plate stretching parameter, the squeeze number, and the bottom-plate suction/injection parameter on the velocity and temperature in the longitudinal direction is studied. The results show that by increasing the squeeze number, the intensity of magnetic field and the magnetic inclined angle velocity near up and bottom plates in the longitudinal direction may enhance the velocity near up and bottom plates however, the velocity near the central region between up and bottom plates decreases. By enhancing the intensity of the magnetic field and the magnetic inclined angle, the temperature rises however, by increasing the squeeze number the temperature decreases. As an intensive conclusion, using the FTADM leads to more accurate results with a smaller number of calculations in comparison with the previous simulations. This may be due to the fact that using the FTADM enables us to incorporate all boundary conditions, however, in the previous semi-analytical methods only a part of boundary conditions may be imposed into the solution.

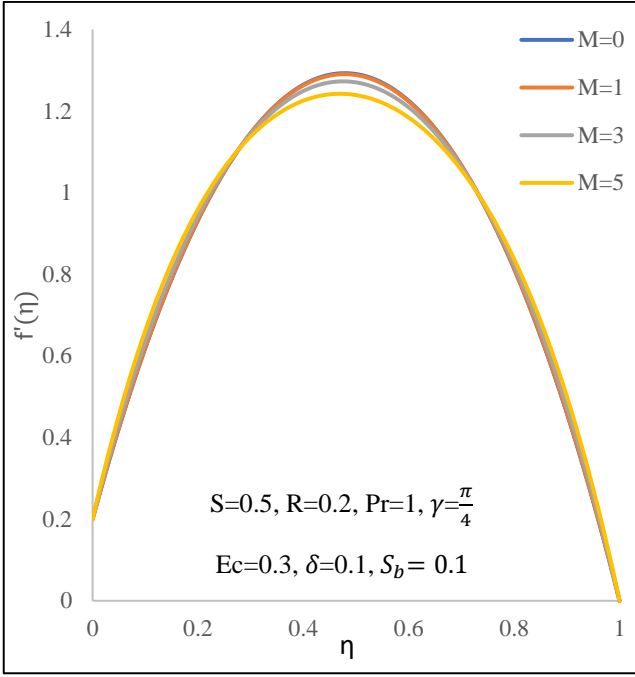 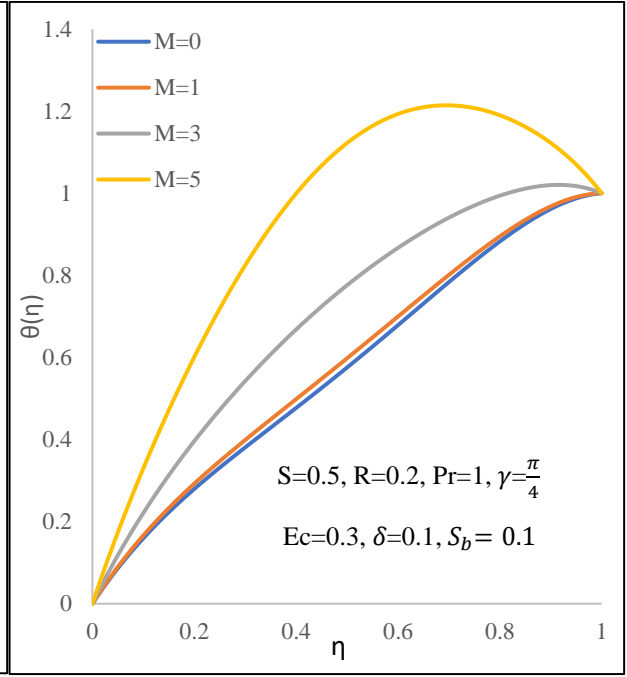

Fig. 4: shows the effect of the intensity of magnetic field M on the velocity profile in the longitudinal direction

Fig. 5: shows the effect of the intensity of magnetic field M on the temperature profile in the longitudinal direction

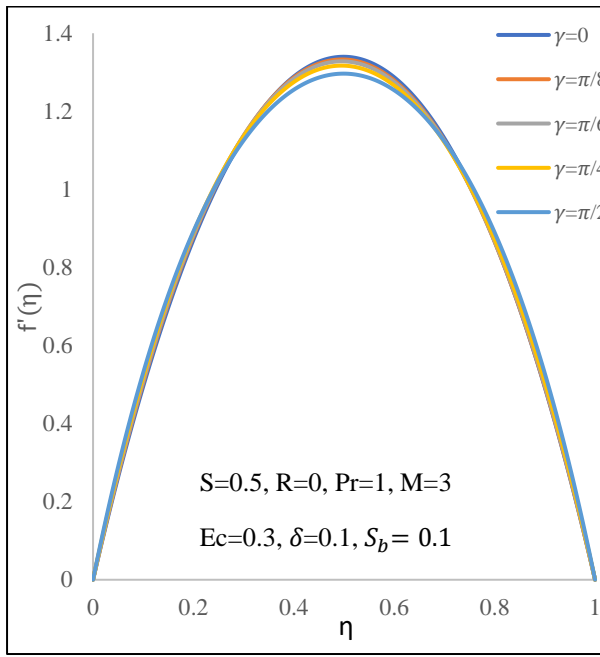

Fig. 6: shows the effect of the inclined angle $\gamma$ on velocity profile in the longitudinal direction

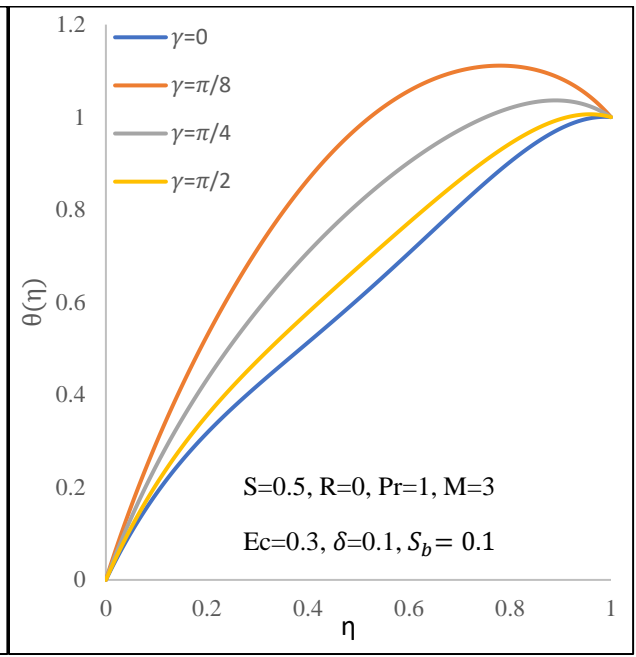

Fig. 7: shows the effect of the magnetic inclination $\gamma$ on the temperature profile in the longitudinal direction

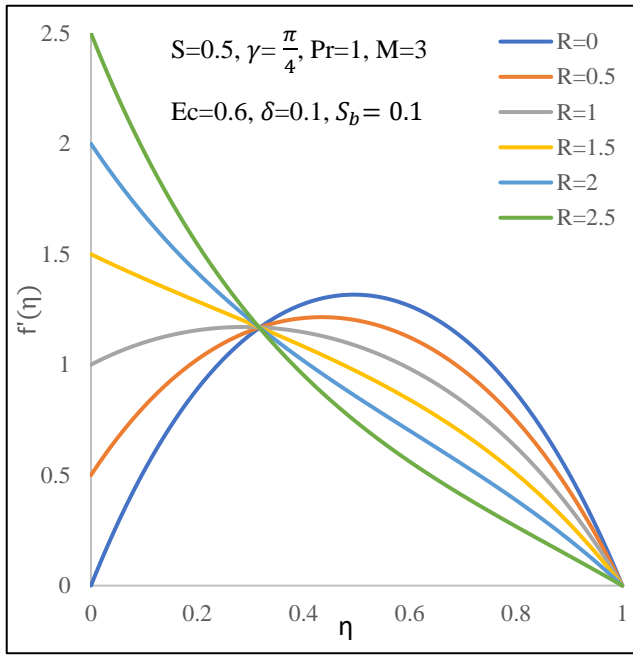

Fig. 8: shows the effect of the lower-plate stretching parameter R on velocity profile in the longitudinal direction

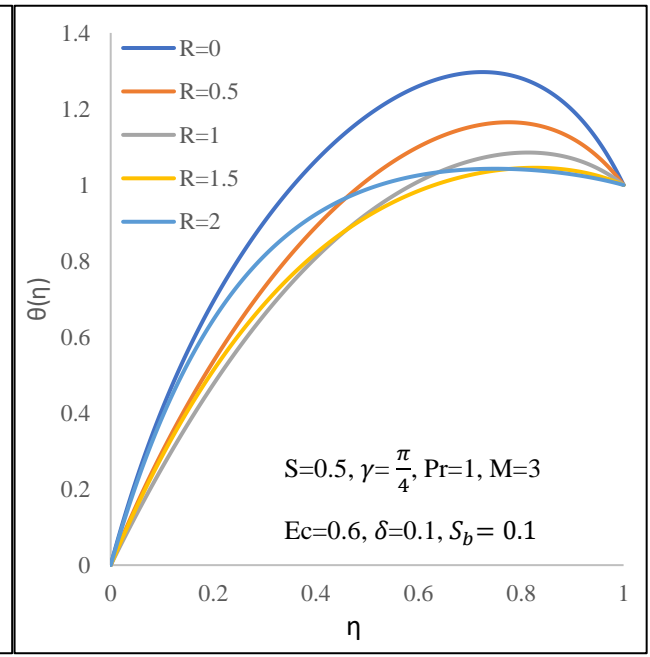

Fig. 9: shows the effect of the lower-plate stretching parameter R on the temperature profile in the longitudinal direction

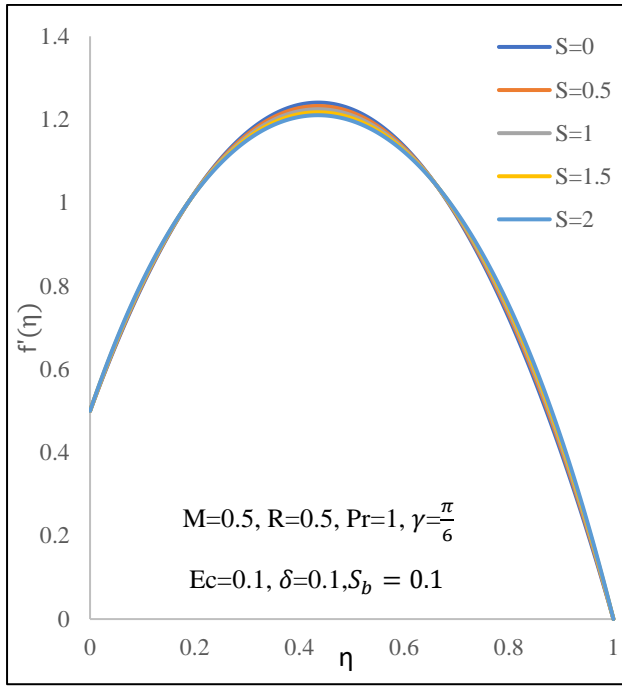

Fig. 10: shows the effect of the squeeze number S on the velocity profile in the longitudinal direction

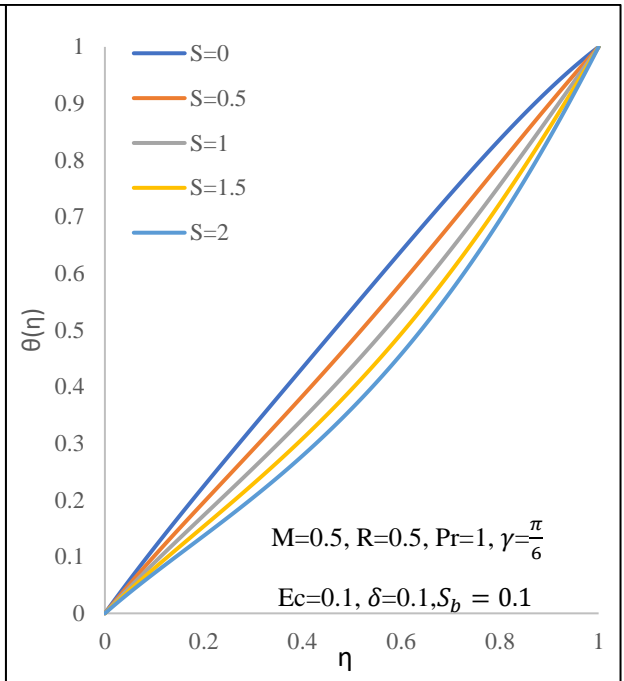

Fig. 11: shows the effect of the squeeze number S on the temperature profile in the longitudinal direction

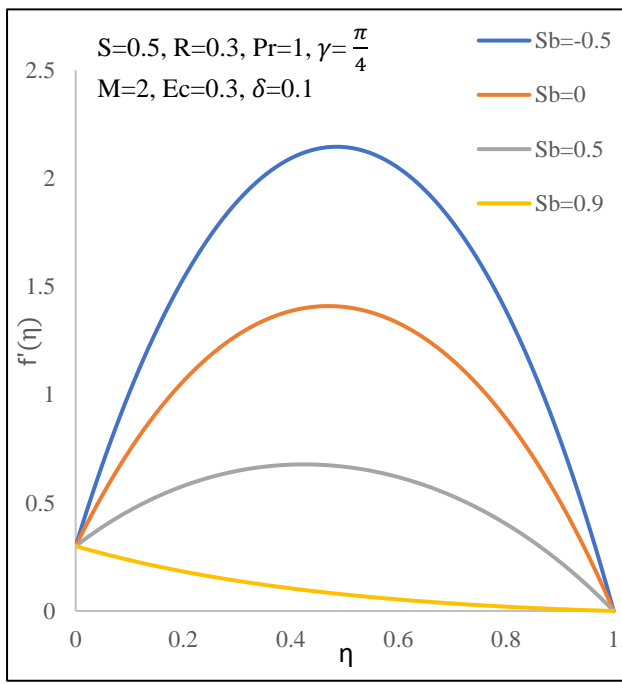

Fig. 12: shows the effect of the lower-plate suction/injection parameter $S_b$ on the velocity profile in the longitudinal direction

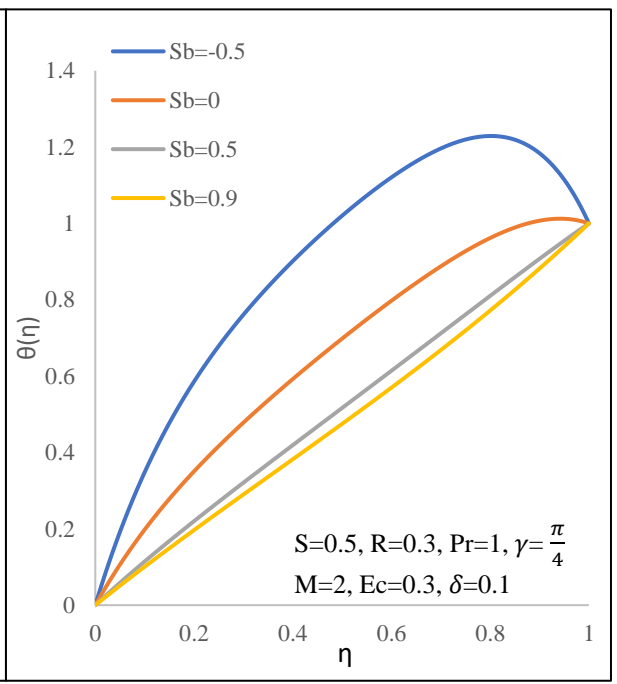

Fig. 13: shows the effect of the lower-plate suction/injection parameter $S_b$ on the on the temperature profile in the longitudinal direction

# Appendix

**Momentum equation:**

Taking the fourth order integral from Eq. (23):

$$\int_0^1\int_0^\eta\int_0^\eta\int_0^\eta \left(f^{iv} - S(\eta f''' + 3f'' + ff'' - ff''') - M^2\sin\gamma(\sin\gamma f'' + 2\delta\cos\gamma f')\right)d\eta\, d\eta\, d\eta\, d\eta = 0, \qquad (60)$$

$$1 - S_b - R - \frac{1}{2}f''(0) - \frac{1}{6}f'''(0) - S\left(I_1 + 3(I_2 - \frac{1}{2}S_b - \frac{1}{6}R) + I_3\right)$$

$$-M^2\sin\gamma\left(\sin\gamma(I_2 - \frac{1}{2}S_b - \frac{1}{6}R) + 2\delta\cos\gamma(I_4 - \frac{1}{6}S_b)\right) = 0, \qquad (61)$$

$$I_1 = \int_0^1\int_0^\eta\int_0^\eta\int_0^\eta \eta f'''(\eta)\, d\eta\, d\eta\, d\eta\, d\eta, \qquad (62)$$

$$I_2 = \int_0^1\int_0^\eta f(\eta)\, d\eta\, d\eta, \qquad (63)$$

$$I_3 = \int_0^1\int_0^\eta\int_0^\eta\int_0^\eta (ff'' - ff''')\, d\eta\, d\eta\, d\eta\, d\eta, \qquad (64)$$

$$I_4 = \int_0^1\int_0^\eta\int_0^\eta f(\eta)\, d\eta\, d\eta\, d\eta, \qquad (65)$$

taking triple integral from Eq. (23):

$$\int_0^1 \int_0^\eta \int_0^\eta (f^{iv} - S(\eta f''' + 3f'' + ff'' - ff''') - M^2 \sin\gamma(\sin\gamma f'' + 2\delta\cos\gamma f'))d\eta d\eta d\eta = 0, \quad (66)$$

$$-R - f''(0) - \frac{1}{2}f'''(0) - S\left(I_{11} + 3(I_{22} - S_b - \frac{1}{2}R) + I_{33}\right)$$

$$-M^2\sin\gamma\left(\sin\gamma(I_{22} - S_b - \frac{1}{2}R) + 2\delta\cos\gamma(I_{44} - \frac{1}{2}S_b)\right) = 0, \quad (67)$$

$$I_{11} = \int_0^1 \int_0^\eta \int_0^\eta \eta f'''(\eta)d\eta d\eta d\eta, \quad (68)$$

$$I_{22} = \int_0^1 f(\eta)d\eta, \quad (69)$$

$$I_{33} = \int_0^1 \int_0^\eta \int_0^\eta (ff'' - ff''')d\eta d\eta d\eta, \quad (70)$$

$$I_{44} = \int_0^1 \int_0^\eta f(\eta)d\eta d\eta, \quad (71)$$

considering $C_1, C_2, C_{11}, C_{22}$ as:

$$C_1 = 1 - S_b - R, \quad (72)$$

$$C_2 = -S\left(I_1 + 3(I_2 - \frac{1}{2}S_b - \frac{1}{6}R) + I_3\right) - M^2\sin\gamma[\sin\gamma(I_2 - \frac{1}{2}S_b - \frac{1}{6}R) + 2\delta\cos\gamma(I_4 - \frac{1}{6}S_b)], \quad (73)$$

$$C_{11} = -R, \quad (74)$$

$$C_{22} = -S\left(I_{11} + 3(I_{22} - S_b - \frac{1}{2}R) + I_{33}\right) - M^2\sin\gamma[\sin\gamma(I_{22} - S_b - \frac{1}{2}R) + 2\delta\cos\gamma(I_{44} - \frac{1}{2}S_b)], \quad (75)$$

substituting Eqs. (72)-(75) into Eqs. (61) and (67):

$$-\frac{1}{2}f''(0) - \frac{1}{6}f'''(0) + C_1 + C_2 = 0, \quad (76)$$

$$-f''(0) - \frac{1}{2}f'''(0) + C_{11} + C_{22} = 0, \quad (77)$$

$$f''(0) = -2(C_{11} + C_{22}) + 6(C_1 + C_2), \quad (78)$$

$$f'''(0) = 6(C_{11} + C_{22}) - 12(C_1 + C_2). \quad (79)$$

Taking Fourier transform of Eq. (23):

$$F\left(f^{iv} - S(\eta f''' + 3f'' + ff'' - ff''') - M^2 \sin\gamma(\sin\gamma f'' + 2\delta\cos\gamma f')\right) = 0, \tag{80}$$

$$F(f'(\eta)) = -f(0) + i\omega F(f(\eta)), \tag{81}$$

$$F(f''(\eta)) = -f'(0) - i\omega f(0) - \omega^2 F(f(\eta)), \tag{82}$$

$$F(f'''(\eta)) = -f''(0) - i\omega f'(0) + \omega^2 f(0) - i\omega^3 F(f(\eta)), \tag{83}$$

$$F\left(f^{iv}(\eta)\right) = -f'''(0) - i\omega f''(0) + \omega^2 f'(0) + i\omega^3 f(0) + \omega^4 F(f(\eta)), \tag{84}$$

substituting Eqs. (81)-(82) into Eq. (80):

$$f = \frac{1}{\omega^4}f'''(0) + \frac{i}{\omega^3}f''(0) - \frac{R}{\omega^2} - \frac{i}{\omega}S_b + \frac{S}{\omega^4}\left(\eta f''' + 3(-R - i\omega S_b - \omega^2 f) + (ff'' + ff''')\right)$$
$$+ \frac{M^2\sin\gamma}{\omega^4}\left(\sin\gamma(-R - i\omega S_b - \omega^2 f) + 2\cos\gamma(-S_b + i\omega f)\right), \tag{85}$$

substituting Eqs. (78) and (79) in Eq. (86):

$$f = \frac{1}{\omega^4}\left(6(C_{11} + C_{22}) - 12(C_1 + C_2)\right) + \frac{i}{\omega^3}\left(-2(C_{11} + C_{22}) + 6(C_1 + C_2)\right) + \frac{S}{\omega^4}\left(\eta f''' + 3(-R - i\omega S_b)\right)$$
$$\frac{S}{\omega^4}\left(-\omega^2 f\right) + \frac{S}{\omega^4}(ff'' + ff''') + \frac{M^2\sin\gamma}{\omega^4}\left(\sin\gamma(-R - i\omega S_b - \omega^2 f) + 2\cos\gamma(-S_b + i\omega f)\right), \tag{86}$$

$$f = \left(-\frac{12}{\omega^4} + \frac{6i}{\omega^3}\right)\left(3S\left(\frac{1}{2}S_b + \frac{1}{6}R\right) + M^2\sin\gamma\left(\sin\gamma(\frac{1}{2}S_b + \frac{1}{6}R) + 2\delta\cos\gamma(\frac{1}{6}S_b)\right)\right)$$
$$+\left(\frac{6}{\omega^4} - \frac{2i}{\omega^3}\right)\left(3S(S_b + \frac{1}{2}R) + M^2 \sin\gamma\left(\sin\gamma(S_b + \frac{1}{2}R) + 2\delta\cos\gamma S_b\right)\right) - R - \frac{3S}{\omega^4}(R + i\omega)$$
$$-\frac{M^2\sin\gamma}{\omega^4}\left(\sin\gamma(R + i\omega S_b) + 2\delta\cos\gamma S_b\right) - \frac{i}{\omega}S_b - \frac{R}{\omega^2} + \left(-\frac{12}{\omega^4} + \frac{6i}{\omega^3}\right)\left(-S(I_1 + 3I_2 + I_3)\right)$$
$$+\left(-\frac{12}{\omega^4} + \frac{6i}{\omega^3}\right)\left(-M^2\sin\gamma(\sin\gamma I_2 + 2\delta\cos\gamma I_4)\right) + \left(\frac{6}{\omega^4} - \frac{2i}{\omega^3}\right)\left(-S(I_{11} + 3I_{22} + I_{33})\right)$$
$$+\left(\frac{6}{\omega^4} - \frac{2i}{\omega^3}\right)\left(-M^2\sin\gamma(\sin\gamma I_{22} + 2\delta\cos\gamma I_{44})\right),$$

$$+\frac{S}{\omega^4}\left(\eta f''' - 3\omega^2 f + (ff'' + ff''')\right) + \frac{M^2\sin\gamma}{\omega^4}\left(-\omega^2\sin\gamma f + 2i\omega\delta\cos\gamma f\right), \tag{87}$$

$$f(\eta) = \sum_{n=0}^{\infty}f_n(\eta), \quad F(\sum_{n=0}^{\infty}f_n(\eta)) = \sum_{n=0}^{\infty}f_n(\omega), \tag{88}$$

$$\eta f''(\eta) = \sum_{n=0}^{\infty}A_n(\eta), \quad F(\sum_{n=0}^{\infty}A_n(\eta)) = \sum_{n=0}^{\infty}A_n(\omega), \tag{89}$$

$$ff''+ff''' = \sum_{n=0}^{\infty} B_n(\eta), \quad F(\sum_{n=0}^{\infty} B_n(\eta)) = \sum_{n=0}^{\infty} B_n(\omega), \tag{90}$$

Making use of Eqs. (88)-(90), Eq. (87) may be rewritten as:

$$\begin{aligned}\sum_{n=0}^{\infty} f_n &= f_0 + (-\frac{12}{\omega^4} + \frac{6i}{\omega^3})\left(-S\left(\int_0^1\int_0^\eta\int_0^\eta\int_0^\eta \sum_{n=0}^{\infty} A_n d\eta d\eta d\eta d\eta + 3\int_0^1\int_0^\eta \sum_{n=0}^{\infty} f_n d\eta d\eta\right)\right)\\ &+(-\frac{12}{\omega^4}+\frac{6i}{\omega^3})\left(\int_0^1\int_0^\eta\int_0^\eta\int_0^\eta \sum_{n=0}^{\infty} B_n d\eta d\eta d\eta d\eta\right) + (-\frac{12}{\omega^4}+\frac{6i}{\omega^3})\left(-M^2 \sin\gamma(\sin\gamma\int_0^1\int_0^\eta \sum_{n=0}^{\infty} f_n d\eta d\eta)\right)\\ &+(-\frac{12}{\omega^4}+\frac{6i}{\omega^3})\left(2\delta\cos\gamma\int_0^1\int_0^\eta\int_0^\eta \sum_{n=0}^{\infty} f_n d\eta d\eta d\eta\right)+(\frac{6}{\omega^4}-\frac{2i}{\omega^3})\left(-S\int_0^1\int_0^\eta\int_0^\eta \widehat{\sum_{n=0}^{\infty} A_n} d\eta d\eta d\eta\right)\\ &+(\frac{6}{\omega^4}-\frac{2i}{\omega^3})\left(-S(3\int_0^1 \widehat{\sum_{n=0}^{\infty} f_n} d\eta + \int_0^1\int_0^\eta\int_0^\eta \widehat{\sum_{n=0}^{\infty} B_n} d\eta d\eta d\eta)\right)+(\frac{6}{\omega^4}-\frac{2i}{\omega^3})\left(-M^2 \sin\gamma(\sin\gamma\int_0^1 \widehat{\sum_{n=0}^{\infty} f_n} d\eta)\right)\\ &+(\frac{6}{\omega^4}-\frac{2i}{\omega^3})\left(2\delta\cos\gamma\int_0^1\int_0^\eta \widehat{\sum_{n=0}^{\infty} f_n} d\eta d\eta\right)+\frac{S}{\omega^4}\left(\widehat{\sum_{n=0}^{\infty} A_n} - 3\omega^2 \widehat{\sum_{n=0}^{\infty} f_n} + \widehat{\sum_{n=0}^{\infty} B_n}\right)\\ &+\frac{M^2 \sin\gamma}{\omega^4}\left(-\omega^2 \sin\gamma \widehat{\sum_{n=0}^{\infty} f_n} + 2i\omega\delta\cos\gamma \widehat{\sum_{n=0}^{\infty} f_n}\right),\end{aligned} \tag{91}$$

where $f_0$ may be written as:

$$\begin{aligned}f_0 &= (-\frac{12}{\omega^4}+\frac{6i}{\omega^3})\left(3S(\frac{1}{2}S_b+\frac{1}{6}R)+M^2\sin\gamma\left(\sin\gamma(\frac{1}{2}S_b+\frac{1}{6}R)+2\delta\cos\gamma(\frac{1}{6}S_b)\right)\right)\\ &(\frac{6}{\omega^4}-\frac{2i}{\omega^3})\left(3S(S_b+\frac{1}{2}R)+M^2\sin\gamma\left(\sin\gamma(S_b+\frac{1}{2}R)+\delta\cos\gamma(S_b)\right)\right)-R-\frac{3S}{\omega^4}(R+i\omega)\\ &-\frac{M^2\sin\gamma}{\omega^4}\left(\sin\gamma(R+i\omega S_b)+2\delta\cos\gamma S_b\right)-\frac{i}{\omega}S_b-\frac{R}{\omega^2},\end{aligned} \tag{92}$$

the recursive equations for momentum equation may be written as:

$$f_0 = F^{-1}(f_0), \tag{93}$$

$$f_1 = (-\frac{12}{\omega^4} + \frac{6i}{\omega^3})\left(-S\left(\int_0^1\int_0^\eta\int_0^\eta\int_0^\eta A_0 d\eta d\eta d\eta d\eta + 3\int_0^1\int_0^\eta f_0 d\eta d\eta\right)\right)$$

$$+(-\frac{12}{\omega^4} + \frac{6i}{\omega^3})\left(\int_0^1\int_0^\eta\int_0^\eta\int_0^\eta B_0 d\eta d\eta d\eta d\eta\right) + (-\frac{12}{\omega^4} + \frac{6i}{\omega^3})\left(-M^2 \sin\gamma(\sin\gamma\int_0^1\int_0^\eta f_0 d\eta d\eta)\right)$$

$$+(-\frac{12}{\omega^4} + \frac{6i}{\omega^3})\left(2\delta\cos\gamma\int_0^1\int_0^\eta\int_0^\eta f_0 d\eta d\eta d\eta\right) + (\frac{6}{\omega^4} - \frac{2i}{\omega^3})\left(-S\int_0^1\int_0^\eta\int_0^\eta A_0 d\eta d\eta d\eta\right)$$

$$(\frac{6}{\omega^4} - \frac{2i}{\omega^3})\left(-S(3\widehat{\int_0^1 f_0 d\eta} + \widehat{\int_0^1\int_0^\eta\int_0^\eta \sum_{n=0}^\infty B_0 d\eta d\eta d\eta})\right) + (\frac{6}{\omega^4} - \frac{2i}{\omega^3})\left(-M^2\sin\gamma(\sin\gamma\widehat{\int_0^1 f_0 d\eta})\right) \quad (94)$$

$$+(\frac{6}{\omega^4} - \frac{2i}{\omega^3})\left(2\delta\cos\gamma\widehat{\int_0^1\int_0^\eta f_0 d\eta d\eta}\right) + \frac{S}{\omega^4}\left(\widehat{A_0 - 3\omega^2 f_0 + B_0}\right)$$

$$+\frac{M^2 \sin\gamma}{\omega^4}\left(-\omega^2 \sin\gamma \widehat{f_0} + 2i\omega\delta\cos\gamma \widehat{f_0}\right)$$

$$f_1 = F^{-1}(\widehat{f_1}),$$

$$f_2 = (-\frac{12}{\omega^4} + \frac{6i}{\omega^3})\left(-S\left(\int_0^1\int_0^\eta\int_0^\eta\int_0^\eta A_1 d\eta d\eta d\eta d\eta + 3\int_0^1\int_0^\eta f_1 d\eta d\eta\right)\right)$$

$$+(-\frac{12}{\omega^4} + \frac{6i}{\omega^3})\left(\int_0^1\int_0^\eta\int_0^\eta\int_0^\eta B_1 d\eta d\eta d\eta d\eta\right) + (-\frac{12}{\omega^4} + \frac{6i}{\omega^3})\left(-M^2 \sin\gamma(\sin\gamma\int_0^1\int_0^\eta f_1 d\eta d\eta)\right)$$

$$+(-\frac{12}{\omega^4} + \frac{6i}{\omega^3})\left(2\delta\cos\gamma\int_0^1\int_0^\eta\int_0^\eta f_1 d\eta d\eta d\eta\right) + (\frac{6}{\omega^4} - \frac{2i}{\omega^3})\left(-S\int_0^1\int_0^\eta\int_0^\eta A_1 d\eta d\eta d\eta\right)$$

$$(\frac{6}{\omega^4} - \frac{2i}{\omega^3})\left(-S(3\widehat{\int_0^1 f_1 d\eta} + \widehat{\int_0^1\int_0^\eta\int_0^\eta \sum_{n=0}^\infty B_1 d\eta d\eta d\eta})\right) + (\frac{6}{\omega^4} - \frac{2i}{\omega^3})\left(-M^2\sin\gamma(\sin\gamma\widehat{\int_0^1 f_1 d\eta})\right) \quad (95)$$

$$+(\frac{6}{\omega^4} - \frac{2i}{\omega^3})\left(2\delta\cos\gamma\widehat{\int_0^1\int_0^\eta f_1 d\eta d\eta}\right) + \frac{S}{\omega^4}\left(\widehat{A_1 - 3\omega^2 f_1 + B_1}\right)$$

$$+\frac{M^2 \sin\gamma}{\omega^4}\left(-\omega^2 \sin\gamma \widehat{f_1} + 2i\omega\delta\cos\gamma \widehat{f_1}\right)$$

$$f_2 = F^{-1}(\widehat{f_2}),$$

and so on

.

.

.

$$f_k = (-\frac{12}{\omega^4} + \frac{6i}{\omega^3})\left(-S\left(\int_0^1\int_0^\eta\int_0^\eta\int_0^\eta A_{k-1}d\eta d\eta d\eta d\eta + 3\int_0^1\int_0^\eta f_{k-1}d\eta d\eta\right)\right)$$

$$+(-\frac{12}{\omega^4} + \frac{6i}{\omega^3})\left(\int_0^1\int_0^\eta\int_0^\eta\int_0^\eta B_{k-1}d\eta d\eta d\eta d\eta\right) + (-\frac{12}{\omega^4} + \frac{6i}{\omega^3})\left(-M^2\sin\gamma(\sin\gamma\int_0^1\int_0^\eta f_{k-1}d\eta d\eta)\right)$$

$$+(-\frac{12}{\omega^4} + \frac{6i}{\omega^3})\left(2\delta\cos\gamma\int_0^1\int_0^\eta\int_0^\eta f_{k-1}d\eta d\eta d\eta\right) + (\frac{6}{\omega^4} - \frac{2i}{\omega^3})\left(-S\int_0^1\int_0^\eta\int_0^\eta A_{k-1}d\eta d\eta d\eta\right)$$

$$(\frac{6}{\omega^4} - \frac{2i}{\omega^3})\left(-S\left(3\int_0^1 \widehat{f_{k-1}d\eta} + \int_0^1\int_0^\eta\int_0^\eta\sum_{n=0}^\infty \widehat{B_{k-1}d\eta d\eta d\eta}\right)\right) + (\frac{6}{\omega^4} - \frac{2i}{\omega^3})\left(-M^2\sin\gamma(\sin\gamma\int_0^1 \widehat{f_{k-1}d\eta})\right) \quad (96)$$

$$+(\frac{6}{\omega^4} - \frac{2i}{\omega^3})\left(2\delta\cos\gamma\int_0^1\int_0^\eta \widehat{f_{k-1}d\eta d\eta}\right) + \frac{S}{\omega^4}\left(\widehat{A_{k-1} - 3\omega^2 f_{k-1} + B_{k-1}}\right)$$

$$+\frac{M^2\sin\gamma}{\omega^4}\left(-\omega^2\sin\gamma\widehat{f_{k-1}} + 2i\omega\delta\cos\gamma\widehat{f_{k-1}}\right)$$

$$f_k = F^{-1}(\widehat{f_k}),$$

which $A_0, A_1, A_2, ..., A_{k-1}, B_0, B_1, B_2, ..., B_{k-1}$ are the Fourier transforms of the Adomian polynomials.

$$f_0(\eta) = (R + 2S_b - 2)\eta^3 + (3 - 3S_b - 2R)\eta^2 + R\eta + S_b,$$

$$f_1(\eta) = \left(\frac{SR^2}{70} + \frac{2SRS_b}{35} - \frac{2SR}{35} + \frac{2SS_b^2}{35} - \frac{4SS_b}{35} + \frac{2S}{35}\right)\eta^7$$

$$+\left(\frac{\delta\sin(2\gamma)M^2 R}{120} + \frac{\delta\sin(2\gamma)M^2 S_b}{120} + \frac{\delta\sin(2\gamma)M^2}{60} - \frac{SR^2}{15} - \frac{7SRS_b}{30} + \frac{7SR}{30} - \frac{S}{5}\right)\eta^6$$

$$+\left(\frac{\sigma_2}{20} - \frac{SR}{5} - \frac{SS_b}{5} - \frac{S}{10} + \frac{M^2 R}{40} + \frac{M^2 S_b}{20} + \frac{2SR^2}{15} + \frac{3SS_b}{10} - \frac{M^2}{20} + \frac{2SRS_b}{5}\right)\eta^5$$

$$+\left(\frac{3S}{4} - \frac{SR}{4} - \frac{SS_b}{4} - \frac{\sigma_2}{8} - \frac{M^2 R}{12} - \frac{M^2 S_b}{8} - \frac{SR^2}{6} - \frac{SS_b^2}{2} + \frac{M^2}{8} - \frac{SRS_b}{2} + \frac{\sigma_4}{12} + \frac{\sigma_1}{8} + \frac{\sigma_3}{24}\right)\eta^4$$

$$+\left(\begin{array}{c}\dfrac{19SR}{42} - \dfrac{24S}{35} + \dfrac{SS_b}{14} + \dfrac{\sigma_2}{10} + \dfrac{11M^2 R}{120} + \dfrac{M^2 S_b}{10} + \dfrac{9SR^2}{70} + \dfrac{43SS_b^2}{70} \\ -\dfrac{M^2}{10} + \dfrac{47SRS_b}{105} - \dfrac{11M^2 R\cos(2\gamma)}{120} - \dfrac{\sigma_1}{10} - \dfrac{\sigma_2}{12} - \dfrac{\sigma_3}{60} + \dfrac{\sigma_6}{12}\end{array}\right)\eta^3$$

$$+\left(\begin{array}{c}\dfrac{5S}{28} - \dfrac{5SR}{28} + \dfrac{13SS_b}{140} - \dfrac{\sigma_2}{40} - \dfrac{M^2 R}{30} - \dfrac{M^2 S_b}{40} - \dfrac{3SR^2}{70} \\ -\dfrac{19SS_b}{70} + \dfrac{M^2}{40} - \dfrac{6SRS_b}{35} + \dfrac{\sigma_4}{30} + \dfrac{\sigma_1}{40} + \dfrac{\sigma_5}{20} + \dfrac{\sigma_6}{20}\end{array}\right)\eta^2,$$

$$\sigma_1 = M^2 S_b \cos(2\gamma)$$

$$\sigma_2 = M^2 \cos(2\gamma)$$

$$\sigma_3 = M^2 R \delta \sin(2\gamma)$$

$$\sigma_4 = M^2 R \cos(2\gamma)$$

$$\sigma_5 = M^2 \delta \sin(2\gamma)$$

$$\sigma_6 = M^2 S_b \delta \sin(2\gamma). \tag{97}$$